# Solid and Quasi-Solid Electrolytes for Zinc Batteries: Balancing Water Activity, Ion Transport, and Interfaces

Souvik Naskar[1], Jiaqian Qin[2], Eric Jianfeng Cheng[1,3,*]

[1]Advanced Institute for Materials Research, Tohoku University, Sendai 980–8577, Japan

[2]Department of Materials Science, Chulalongkorn University, Bangkok 10330, Thailand

[3]Department of Chemical Engineering, Chulalongkorn University, Bangkok 10330, Thailand

[*]Correspondence: ericonium@tohoku.ac.jp

**Keywords:** zinc-ion batteries; solid and quasi-solid electrolytes; water activity; $Zn^{2+}$ transport; electrode-electrolyte interfaces

## Abstract

Zinc-ion batteries (ZIBs) offer a compelling route to safe and low-cost energy storage, yet their reliance on aqueous electrolytes promotes hydrogen evolution, corrosion, cathode dissolution, and nonuniform zinc (Zn) deposition. Replacing the liquid with a solid electrolyte (SE) or quasi-solid electrolyte (QSE) can suppress these processes, but it also removes the medium that enables rapid $Zn^{2+}$ transport and conformal electrode contact. This tension, the price of removing water, has been obscured by inconsistent use of the term "solid state" and by comparisons based largely on bulk ionic conductivity. Here, we examine Zn electrolytes across a continuum from water-rich hydrogels to dry polymers, solvated crystals, and inorganic conductors. We distinguish water content from thermodynamic water activity and classify these materials according to phase state, mobile-solvent fraction, and dominant transport mechanism. We argue that neither high conductivity nor nominally water-free composition reliably predicts cell performance: electrolyte thickness, $Zn^{2+}$ transference, interfacial resistance, and evolving contact often determine the practical outcome. Controlled-solvation and hybrid electrolytes therefore provide the most credible near-term path, whereas genuinely solvent-free $Zn^{2+}$ conductors remain a longer-term scientific target. Progress will require transparent reporting of solvent state and validation using thin electrolytes, realistic electrode loadings, limited Zn excess, and calendar-life testing.

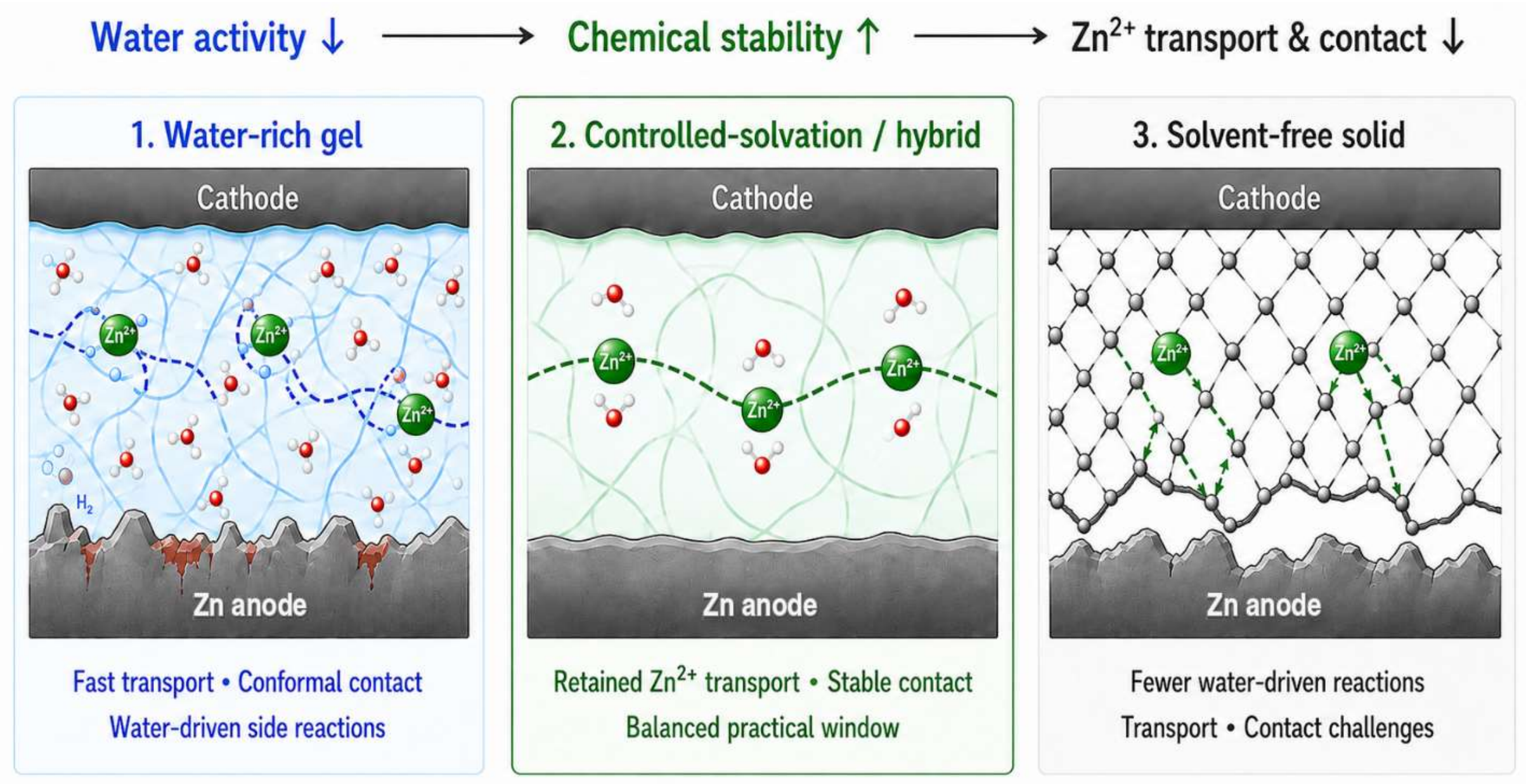

**Graphical Abstract:** A critical perspective on Zn electrolytes spanning hydrogels, controlled-solvation and dry-polymer systems, solvated crystals, and inorganic conductors. Practical battery performance depends on the interplay among solvation state, $Zn^{2+}$ transport, electrolyte thickness, interfacial resistance, and evolving contact: not on water removal or bulk conductivity alone.

## 1. Introduction: Why solidify a Zn electrolyte?

An SE alters Zn chemistry by replacing bulk-liquid transport with ion migration through a solid or polymeric matrix. Solidification can suppress water-driven corrosion, hydrogen evolution, cathode disintegration, electrolyte leakage, and uncontrolled $Zn^{2+}$ concentration gradients and may mechanically regulate Zn deposition [1–9]. It also introduces new constraints, including limited segmental motion, insufficient electrode contact, elevated interfacial resistance, and local current-density heterogeneity [4,5,7–9]. Crucially, a gel containing an immobilized aqueous electrolyte is not necessarily water-free; it may retain almost the same $Zn^{2+}$ hydration chemistry as the corresponding liquid electrolyte. Removing water is particularly difficult for $Zn^{2+}$ because its divalent charge produces high charge density and strong ion–dipole interactions. $Zn^{2+}$ therefore forms a stable hydration shell, commonly approximated as $[Zn(H_2O)_6]^{2+}$ [10,11], and transferring it into a nonaqueous or solid coordination environment entails a substantial desolvation penalty. Water-poor Zn electrolytes must replace water with coordinating polymers, anions, ionic liquids, or other ligands while preserving rapid ligand exchange and ion migration. If coordination is too weak, Zn salts may not dissociate sufficiently; if it is too strong, $Zn^{2+}$ becomes immobilized and conductivity declines. Designing an anhydrous $Zn^{2+}$-conducting SE therefore requires a balance among salt dissociation, $Zn^{2+}$ coordination, ligand exchange, interfacial desolvation, and mechanical stability [12–15].

Compared with lithium-based systems, Zn offers high crustal abundance, relatively low cost, and compatibility with aqueous processing. Zn metal also has theoretical gravimetric and volumetric capacities of 820 mAh $g^{-1}$ and 5855 mAh $cm^{-3}$, respectively. These advantages make zinc batteries attractive for stationary storage, flexible electronics, and wearable devices [16–25]. Nevertheless, practical development remains constrained by reactions at the electrode–electrolyte interfaces. In conventional aqueous electrolytes, Zn metal is prone to dendrite growth, hydrogen evolution, corrosion, passivation, and formation of irreversible zinc hydroxide sulfate by-products. These processes consume active Zn and electrolyte, increase internal resistance, reduce Coulombic efficiency, and eventually cause short circuits or rapid capacity decay [26–34]. The solid-electrolyte interphase (SEI), or more generally the interfacial layer, strongly influences dendrite growth, electrochemical stability, Zn corrosion, and electrolyte decomposition; interface control is therefore central to practical zinc-battery development [35–39].

SEs and QSEs have emerged as promising approaches for mitigating electrolyte leakage and stabilizing Zn electrodes. However, the term “solid-state ZIB” is used inconsistently. Several systems described as solid-state batteries employ hydrogels or gel-polymer electrolytes prepared by immobilizing aqueous Zn-salt solutions within polymer networks; these should be classified as QSE systems [40–42]. Such hydrated gels can provide liquid-like ionic conductivity and excellent mechanical flexibility, but their mobile water fraction leaves them vulnerable to dehydration, freezing, swelling, and water-driven interfacial reactions [42–44].

In contrast, genuinely solid systems contain no continuous liquid phase and may employ solvent-free polymers, plastic-crystalline solids, crystalline frameworks, chalcogenides, ceramics, or other inorganic $Zn^{2+}$ conductors [45–48]. These electrolytes reduce leakage and volatility but often exhibit lower room-temperature $Zn^{2+}$ conductivity and substantial solid–solid interfacial resistance. The distinction matters: QSEs are primarily limited by water management and long-term chemical stability, whereas genuinely solid electrolytes (SEs) are primarily limited by bulk $Zn^{2+}$ transport and electrode–electrolyte contact. This review evaluates SEs and QSEs for zinc batteries through the trade-off between reducing water activity and maintaining rapid $Zn^{2+}$ transport and stable electrode contact. It establishes a composition-and mechanism-based classification, distinguishes water content from thermodynamic activity, and compares bulk conductivity with evolving cell-level resistance. Particular attention is given to $Zn^{2+}$ transport mechanisms, interphase formation, cathode loading, Zn utilization, electrolyte inventory, pouch-cell validation, calendar life, manufacturing, and sustainability. Controlled-solvation and hybrid electrolytes are identified as the most practical near-term route, whereas genuinely solvent-free $Zn^{2+}$ conductors remain a longer-term scientific objective. Figure 1 summarizes this electrolyte continuum.

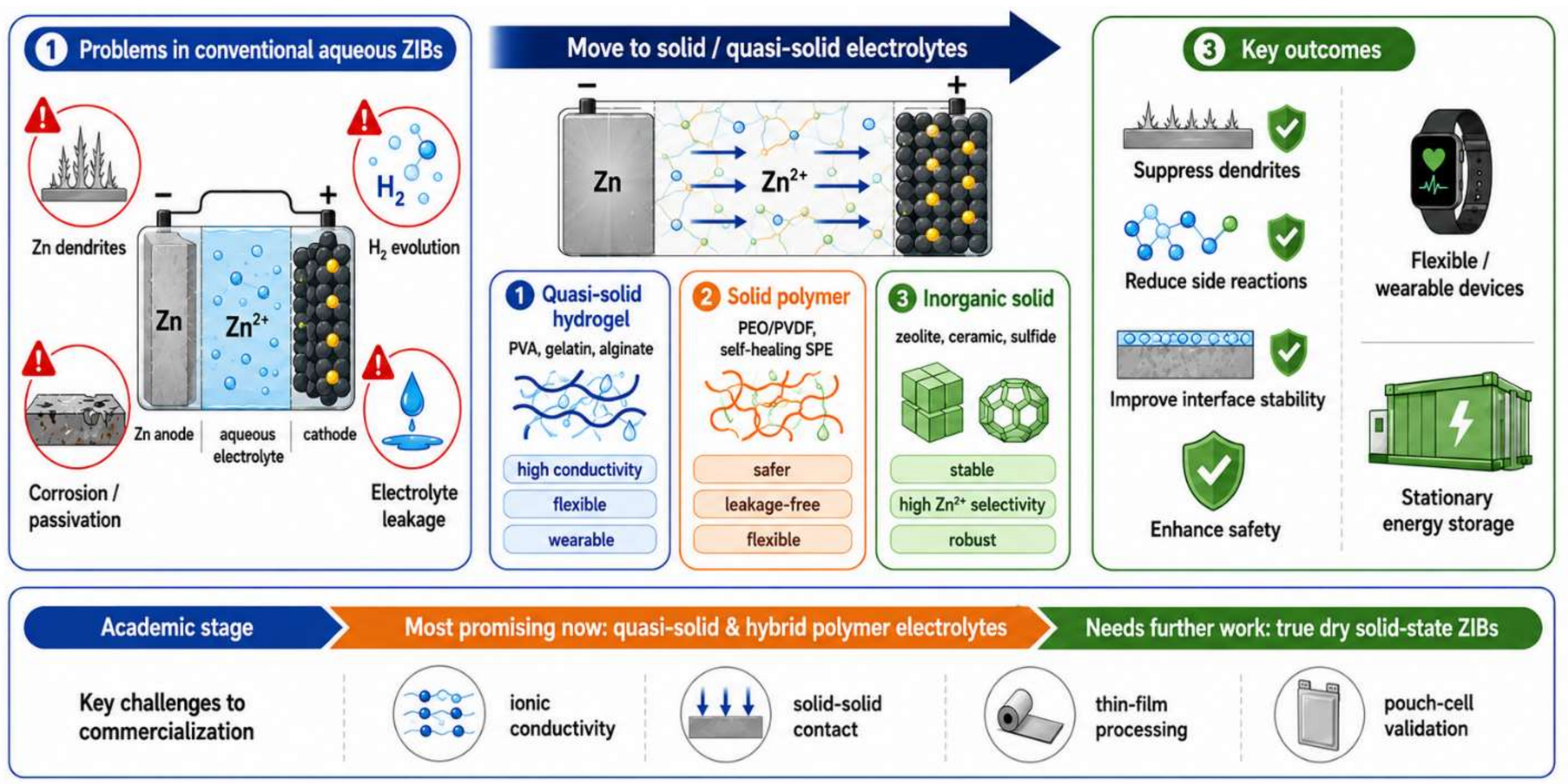

**Figure 1.** Continuum from aqueous electrolytes to quasi-solid and SEs for zinc batteries, highlighting principal benefits and commercialization challenges.

## 2. What counts as solid?

A self-supporting shape, absence of visible leakage, or low solvent content does not by itself establish a true SE. The decisive question is whether $Zn^{2+}$ transport occurs mainly through a solid host or through a remaining mobile liquid phase. Because these categories form a continuum, every study should report the water and solvent mass fractions, residual moisture, salt-to-solvent ratio, drying temperature and duration, relative-humidity storage conditions, phase state, water activity where measurable, and evidence for the presence or absence of a mobile liquid phase. Ionic conductivity should also be interpreted together with temperature dependence, $Zn^{2+}$ transference number, electrolyte thickness, and interfacial resistance. Table 1 classifies electrolytes by solvent content and transport mechanism. Raman, Fourier-transform infrared (FTIR), and nuclear magnetic resonance (NMR) spectroscopy can probe water structure, hydrogen bonding, ion–water coordination, and molecular mobility, thereby distinguishing bulk-like, intermediate, and strongly associated water. However, the resulting "free" and "bound" water fractions are operational descriptors and should not be equated directly with thermodynamic water activity. Because water activity reflects water chemical potential, claims of reduced activity should preferably be supported by direct equilibrium measurements, such as vapor-pressure, relative-humidity-equilibration, osmotic, or electrochemical-potential methods. Spectroscopy, differential scanning calorimetry (DSC), thermal analysis, and molecular simulations should therefore be treated as complementary mechanistic evidence rather than standalone measurements of water activity [49–52].

Zn electrolytes should therefore be classified by phase state, mobile-solvent fraction, water activity, and dominant ion-transport mechanism—not simply by whether a material appears self-supporting.

**Table 1.** Classification of zinc electrolytes by solvent content and dominant transport mechanism.

| Electrolyte class | Mobile solvent | Physical state | Dominant transport mechanism | Recommended designation |
|---|---|---|---|---|
| **Inorganic SE** [48,53–58] | None or trace residual moisture | Fully solid, crystalline or amorphous phase | Migration through lattice sites, vacancies, grain boundaries, or interconnected pores | **True solid-state** |
| **Dry solid-polymer electrolyte (SPE)** [12,15,59] | None or very low | Solid polymer–salt phase without a separate liquid phase | Polymer-segment-assisted hopping between coordinating sites | **True solid-polymer electrolyte** |
| **Solvated solid or solid eutectic** [60–63] | Limited solvent strongly coordinated within one solid phase | Solid or plastic-crystalline phase; no macroscopic liquid | Ligand exchange and hopping through a dynamically disordered coordination network | **Solvated SE** |

| Electrolyte class | Mobile solvent | Physical state | Dominant transport mechanism | Recommended designation |
|---|---|---|---|---|
| **Hydrogel electrolyte** [64–70] | High aqueous-electrolyte fraction | Water-swollen polymer network | Hydrated $Zn^{2+}$ diffusion through an aqueous phase | **Aqueous quasi-SE** |
| **Confined liquid electrolyte [14,71,72]** | High | Liquid held within a porous separator, membrane | Liquid-phase diffusion within confined pores | **Confined-liquid electrolyte**, not a true SE |
| **Liquid electrolyte** [73–79] | Continuous bulk liquid | Flowable liquid | Solvated-ion diffusion and migration | **Liquid electrolyte** |

## 3. How does $Zn^{2+}$ move?

No single mechanism can explain $Zn^{2+}$ transport because the mobile species and its coordination environment vary among liquid [80], gel [67], polymeric [81], eutectic [82], molecular-crystal [60], and inorganic electrolytes [58]. Relevant mechanisms include vehicular diffusion, coordination exchange, polymer-segment-assisted motion, molecular rotation, framework hopping, grain-boundary conduction, and humidity-assisted transport. Several may coexist, with their relative contributions depending on temperature, microstructure, host mobility, salt dissociation, and water or solvent concentration. Equation (1) gives the Nernst–Einstein relation for estimating the contribution of mobile Zn-containing species to ionic conductivity [83,84].

$$\sigma_{Zn} = \frac{n_{Zn} e^2 D_{Zn} Z_{Zn}^2}{k_B T} \quad (1)$$

In Equation (1), the $Zn^{2+}$ conductivity contribution is expressed in S $m^{-1}$; the mobile-carrier number density in $m^{-3}$; and the long-time Zn self-diffusion coefficient in $m^2$ $s^{-1}$. The Zn charge number is +2, e is the elementary charge ($1.602 \times 10^{-19}$ C), k is the Boltzmann constant, and T is absolute temperature (K). This relation is exact only for an idealized independent-particle system. Three complications are especially important for $Zn^{2+}$. (1) Analytical concentration is not necessarily carrier concentration: Zn may exist as solvated ions, contact ion pairs, ligand-bound complexes, or larger clusters, and neutral complexes may diffuse without transporting net charge. (2) The squared charge-number factor is 4, but it does not guarantee high conductivity. Strong coordination to solvents and anions can reduce mobility and slow ligand exchange. (3) Self-diffusion is not charge diffusion. Pulsed-field-gradient NMR or molecular-dynamics simulations measure the self-diffusion of Zn nuclei averaged over their environments, whereas conductivity depends on collective charge displacement. Correlated motion of Zn-containing species and anions can therefore yield appreciable self-diffusion but little net current [76].

**Table 2.** Summary of $Zn^{2+}$ transport mechanisms in zinc electrolytes.

| Mechanism | Mobile Zn species | Molecular description | Principal limitation |
|---|---|---|---|
| Vehicular [85–89] | Solvated or hydrated $Zn^{2+}$ complex | $Zn^{2+}$ moves together with its solvation shell | Viscosity, tortuosity, and large hydrodynamic radius |
| Coordination exchange [90–96] | Partly solvated $Zn^{2+}$ | $Zn^{2+}$ transfers between neighboring O-, N- or S-donor sites | Excessively strong Zn–ligand binding |
| Segmental [13,97–102] | Polymer-coordinated $Zn^{2+}$ | Polymer motion opens and closes coordination sites | Glass transition, crystallinity, and slow chain relaxation |
| Humidity-assisted [14,103] | Hydrated or water-bridged $Zn^{2+}$ | Adsorbed/bound water creates coordination bridges and increases mobility | Humidity dependence and loss of true dry-state behavior |
| Rotational [84,104–107] | $Zn^{2+}$ in molecular solids | Local molecular rotation assists site-to-site transfer | Restricted rotational disorder and ion pairing |
| Framework hopping [48,55,60,108–112] | Partly coordinated or nearly bare $Zn^{2+}$ | $Zn^{2+}$ hops between lattice sites or connected channels | High migration barrier and poor site connectivity |

| Mechanism | Mobile Zn species | Molecular description | Principal limitation |
|---|---|---|---|
| Grain-boundary/interfacial [48,53,55,56,58,113] | $Zn^{2+}$ at disordered interfaces | Transport follows defect-rich boundaries or heterogeneous interfaces | Blocking boundaries, porosity and inconsistent microstructure |

**Vehicular transport**

Vehicular transport dominates aqueous liquids and many hydrogel or gel-polymer electrolytes. In dilute aqueous media, $Zn^{2+}$ is a solvated cation commonly approximated as $[Zn(H_2O)_6]^{2+}$. Because water is mobile and has low viscosity, the coordination shell moves with the ion, producing comparatively high conductivity. Transport slows as viscosity, polymer-network tortuosity, or salt concentration increases, and the hydrated complex is substantially larger than bare $Zn^{2+}$.

Thus, despite being self-supporting, hydrogels can exhibit liquid-like conductivities. For example, a zwitterionic hydrogel reported a $Zn^{2+}$ transference number of 0.81 and an ionic conductivity of 59.0 mS $cm^{-1}$. These values reflect hydrated-ion transport through a water-rich polymer network, not dry solid-state hopping [114].

**Coordination-exchange transport**

$Zn^{2+}$ can migrate without carrying a complete solvent shell if it repeatedly releases one ligand and binds another nearby donor group. This coordination-exchange mechanism may occur between water molecules, polymer functional groups, anions, oligomers, or framework sites. Effective transport requires a delicate balance. If Zn-host coordination is too weak, the Zn salt remains associated with its anion. If it is too strong, $Zn^{2+}$ becomes trapped at a coordination site. Because $Zn^{2+}$ can adopt several coordination numbers and geometries, suitably dynamic O- or N-donor networks can promote exchange, whereas static, strongly coordinating sites may instead immobilize it.

Yang et al. reported a 128 μm sodium-polyacrylate-based QSE with ion-selective pathways. Zn symmetric cells achieved a reported Coulombic efficiency of 99.7% over 2400 cycles and cycled for 600 h at a Zn depth of discharge of 85.6%. A Zn||$V_2O_5$ pouch cell delivered 1.13 Ah at a cathode loading of 31.3 mg cm$^{-2}$ [94].

**Polymer-segment-assisted transport**

In dry solid-polymer electrolytes, $Zn^{2+}$ coordinates with ether oxygen, carbonyl, nitrile, or other donor groups. Thermal motion of the polymer segments continuously rearranges these coordination sites, allowing $Zn^{2+}$ to transfer between neighboring sections of the chain. Transport is therefore strongly coupled with polymer relaxation and usually increases above the glass-transition temperature.

This mechanism is intrinsically difficult for $Zn^{2+}$ because its high charge density produces strong, long-lived interactions with polymer donor groups. Conventional PEO-based solvent-free Zn conductors have room-temperature conductivities of approximately $10^{-7}$ S cm$^{-1}$, partly because slow polymer dynamics support only short-range $Zn^{2+}$ motion. Small-molecule rotational solids increased conductivity to 0.046 mS cm$^{-1}$ by reducing dependence on whole-chain motion [15]. Salt dissociation is another limitation. Increasing the dielectric constant of a PVDF-based host enhanced salt dissociation and produced a conductivity of 1.07 mS cm$^{-1}$, showing that polymer mobility alone is insufficient: the matrix must also separate $Zn^{2+}$ from its counterion [115].

**Humidity or bound water-assisted transport**

Trace or bound water can increase $Zn^{2+}$ conductivity by plasticizing polymers, expanding free volume, weakening Zn-polymer or Zn-anion interactions, and forming water-bridged coordination pathways. Consequently, an electrolyte that appears solid may exhibit conductivity that is strongly dependent on drying history and ambient humidity.

A quantitative example is a polyethylene glycol (PEG)-based viscoelastic electrolyte containing controlled amounts of bound water. Increasing water content from 0 to 19.87 wt% expanded the measured free volume from 152 to 163 Å$^3$; an $H_2O$-to-ether-oxygen ratio of 1:1 yielded approximately three times the Zn-ion diffusion coefficient measured at a 1:3 ratio [83]. Bound water can therefore shorten the distance between accessible coordination environments and facilitate exchange even when no bulk-like free-water phase is claimed. Humidity-assisted conduction is not necessarily undesirable, but it must be distinguished from intrinsic dry-state transport. A separate $ZnPS_3$ study showed that water-vapor exposure can strongly alter ionic conduction [56]. Conductivities measured after ambient storage should therefore not be compared directly with values obtained after vacuum drying or glovebox handling.

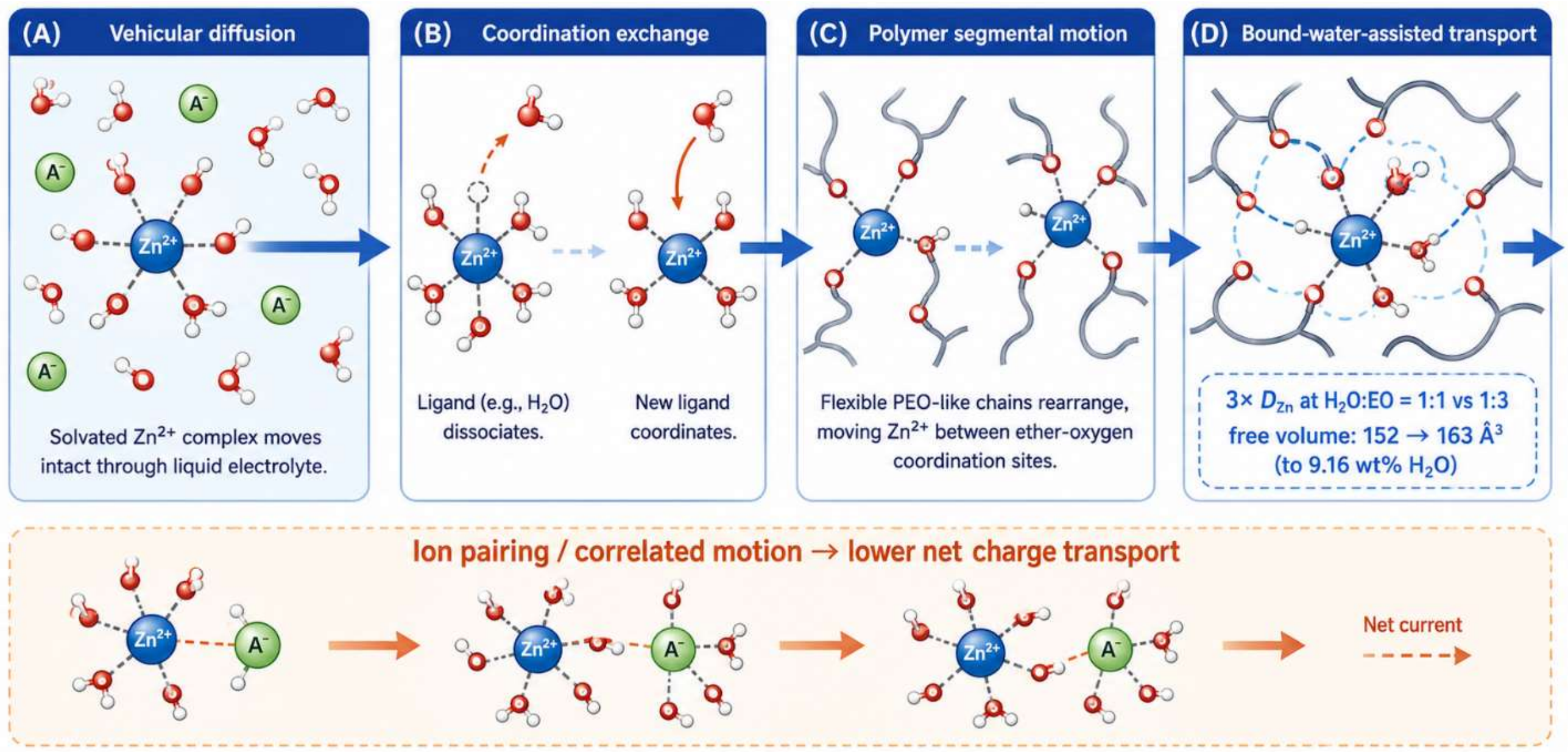

**Figure 2.** Schematic illustration of $Zn^{2+}$ transport in liquids and QSEs through vehicular diffusion, coordination exchange, polymer-segmental motion, and bound-water-assisted migration. Ion pairing and correlated motion can reduce net $Zn^{2+}$ charge transport.

**Rotationally assisted transport**

In molecular or plastic-crystalline solids, local molecular rotations can transiently weaken coordination bonds and reorient donor groups, enabling $Zn^{2+}$ to hop between sites. This is sometimes described as a rotationally assisted or paddle-wheel-like mechanism. Unlike polymer transport, the entire long chain does not need to rearrange.

An ordered zinc bis(trifluoromethanesulfonyl)imide–succinonitrile supramolecular crystal, $Zn(TFSI)_2\cdot 3SN$, achieved 0.62 mS cm$^{-1}$ at 25 °C and 0.0326 mS cm$^{-1}$ at −35 °C, with a reported $Zn^{2+}$ transference number of 0.97. Its ordered three-dimensional tunnels support longer-range transport than conventional segmentally limited polymer hosts [84].

## Framework hopping

In inorganic or crystalline electrolytes, $Zn^{2+}$ moves between vacancies, interstitial sites, or coordination polyhedra. The hopping rate depends on the site-energy difference, bottleneck size, defect concentration, and connectivity of the conduction pathway.

A solid-state $ZnPS_3$ electrolyte was reported to contain interconnected $Zn^{2+}$ migration channels, with a calculated diffusion barrier of approximately 0.3 eV and a conductivity of 2 mS cm$^{-1}$ at 30 °C. However, the authors also identified bound water on the grains, so the high conductivity should not be interpreted as entirely water-independent lattice transport [55]. By comparison, the older $ZnZr_4(PO_4)_6$ framework exhibited only 0.00134 mS cm$^{-1}$ at 500 °C and 1.57 mS cm$^{-1}$ at 900 °C. This temperature requirement illustrates the difficulty of moving a high-charge-density divalent ion through a rigid oxide framework [15].

## Grain-boundary and interfacial transport

Grain boundaries can either accelerate or obstruct $Zn^{2+}$ transport. Structurally disordered boundaries may contain greater free volume, weaker coordination, and more defects than the crystalline bulk, thereby providing low-energy pathways. Conversely, poorly connected grains, insulating secondary phases, and intergranular voids can increase resistance.

In crystallized eutectic Zn conductors, heterogeneous interfaces generated $Zn^{2+}$-percolating pathways and produced a conductivity of 0.0378 mS cm$^{-1}$ at 30 °C, with a reported transference number of 0.64. Here, transport is enhanced by interfacial networks rather than the homogeneous crystalline bulk [60]. Similarly, confining a deep-eutectic electrolyte within PCN-222 metal–organic framework channels produced a conductivity of 0.313 mS cm$^{-1}$, an activation energy of 0.12 eV, and a reported $Zn^{2+}$ transference number of 0.74. Transport occurs through a hybrid network containing a confined mobile phase and framework interfaces, not through a completely dry MOF lattice [116].

## Ion pairing and charge-density limitations

The principal chemical limitation is the high charge density of $Zn^{2+}$. Compared with monovalent ions, $Zn^{2+}$ interacts more strongly with solvents, polymer donor groups, and counterions. This produces three competing consequences: (1) strong solvation improves salt dissolution but increases the size of the transported complex; (2) strong host coordination promotes salt dissociation but can immobilize $Zn^{2+}$;

and (3) strong anion association forms contact ion pairs or aggregates that lower the effective concentration of mobile $Zn^{2+}$ carriers.

At low salt loading, too few carriers are available. At high loading, neutral or weakly charged ion pairs and aggregates become more abundant, viscosity rises, and the coordination network may become saturated. Conductivity can therefore pass through a maximum rather than increase monotonically with Zn-salt concentration. This behavior was observed in the molecule-flexible-solid study: strong cation–anion pairing limited $Zn^{2+}$ mobility above an optimum salt concentration, whereas weaker association extended conductivity toward 0.1 mS $cm^{-1}$ [15]. Total ionic conductivity is also not specific to $Zn^{2+}$. Fast anion, proton, or solvent-associated transport may produce a high measured conductivity while $Zn^{2+}$ remains comparatively slow. Conductivity should therefore be reported together with $Zn^{2+}$ transference number, self-diffusion coefficients, activation energy, salt speciation, and spectroscopic or computational analysis of the coordination environment. DC-polarization-derived transference numbers require particular caution because interfacial reactions, concentration polarization, and changes in Zn–electrolyte contact can distort the apparent steady-state current. Figure 3 summarizes $Zn^{2+}$ transport in dry polymer and crystalline SEs and the associated microstructural limitations.

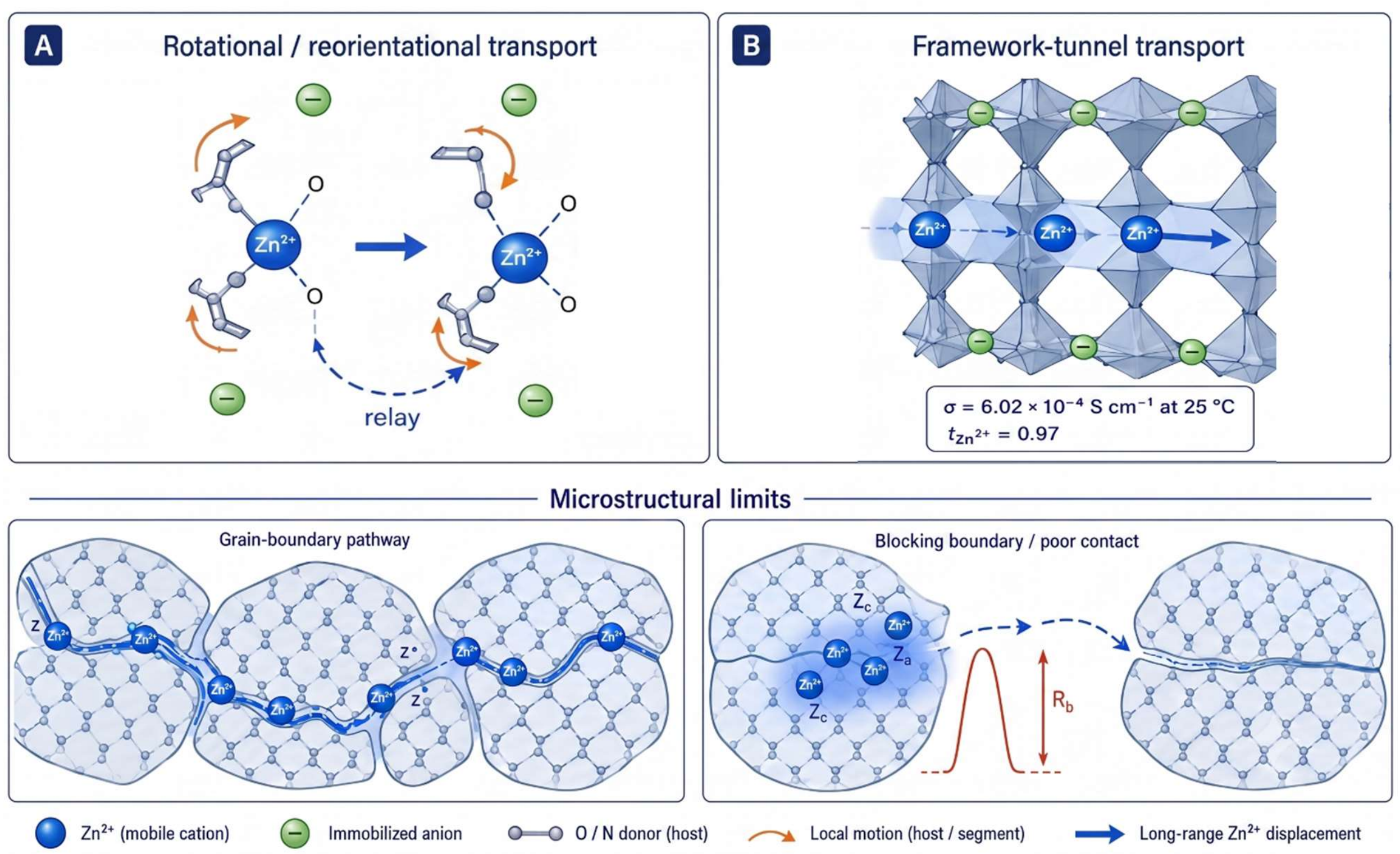

**Figure 3.** Schematic illustration of $Zn^{2+}$ transport in dry polymer and crystalline SEs through molecular reorientation and framework-tunnel conduction. Grain boundaries may facilitate or impede long-range transport, whereas poor interfacial contact interrupts $Zn^{2+}$ pathways.

## 4. Electrolyte designs across the solvent-activity continuum

### Water-rich and confined-solvent polymers

Hydrogel and gel electrolytes are the most mature QSEs for flexible zinc batteries. Their principal strength is that a polymer network immobilizes an aqueous electrolyte without fully removing it, preserving high ionic conductivity while improving leakage resistance and mechanical integrity. A representative example is a hierarchical gelatin-grafted polyacrylamide/polyacrylonitrile (gelatin-g-PAM/PAN) electrolyte, in which a hydrogel is confined within an electrospun PAN membrane. The resulting flexible battery combined high capacity, areal energy density, cycling stability, and resistance to cutting, bending, hammering, puncturing, washing, and fire. It continued to power an electronic watch after repeated cutting and retained an open-circuit voltage of 1.43 V, or 95.9% of its initial value (Figure 4a,b) [20]. Self-healing hydrogels extend this concept to deformable electronics. In a PVA–Zn/Mn hydrogel, reversible hydrogen bonding enables healing after mechanical damage. Combined with all-in-one $VS_2$/carbon-cloth cathodes and Zn/carbon-cloth anodes, the device remained functional after cutting, healing, and bending. At 50 mA $g^{-1}$, the initial capacities in flat, 60°, 90°, and 180° configurations were 165, 180, 169, and 156 mAh $g^{-1}$, respectively; after 30 cycles, 61–75% of the capacity remained (Figure 4c–e) [117]. The larger losses at greater bending angles were attributed to compressive and shear stresses that weakened interfacial contact. Recent work has therefore shifted from simple hydrogel immobilization toward deliberate interfacial regulation.

The sodium-alginate/electron–ion dual-channel (SA/EIDC) gel is one example. Sodium alginate alters the $Zn^{2+}$ solvation environment and reduces active water, whereas poly(3,4-ethylenedioxythiophene):polystyrene sulfonate (PEDOT:PSS) layers provide ion-transport pathways through PSS–$SO_3^-$ groups and electronic pathways through the PEDOT π-conjugated network. The SA gel alone partially regulates hydrated $Zn^{2+}$ through –$COO^-$ groups, but its disordered chains impede $Zn^{2+}$ transport and create an uneven interfacial $Zn^{2+}$ distribution (Figure 4f). Adding the EIDC layer accelerates $Zn^{2+}$ migration and deposition, reduces polarization, and suppresses corrosion. Charge–discharge profiles at the first, 2000th, 3000th, and 4000th cycles demonstrate long-term stability (Figure 4g). A Zn||$MnO_2$ pouch cell with the EIDC layer positioned only between the SA gel and Zn anode also retained nearly constant capacity at different bending angles (Figure 4h) [118].

Hydrogels remain water-rich and require durable encapsulation and water retention to limit evaporation, freezing, microbial degradation, and gas generation.

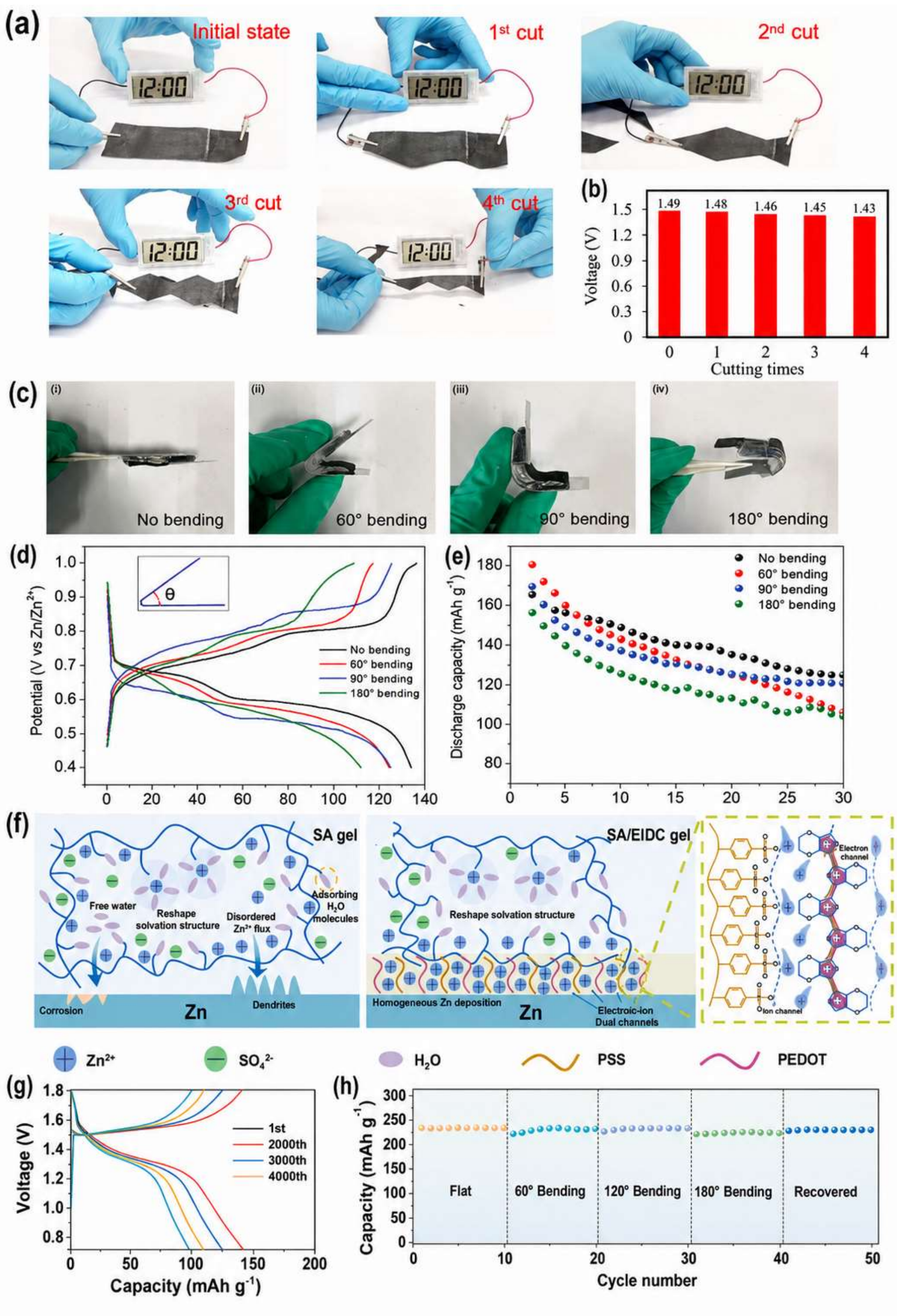

(a)
Initial state
1st cut
2nd cut
3rd cut
4th cut
(b)
Voltage (V)
Cutting times
1.49
1.48
1.46
1.45
1.43
(c)
No bending
60° bending
90° bending
180° bending
(d)
Potential (V vs Zn/Zn2+)
No bending
60° bending
90° bending
180° bending
(e)
Discharge capacity (mAh g-1)
No bending
60° bending
90° bending
180° bending
(f)
SA gel
Free water
Reshape solvation structure
Disordered Zn2+ flux
Adsorbing H2O molecules
Corrosion
Zn
Dendrites
SA/EIDC gel
Reshape solvation structure
Homogeneous Zn deposition
Electroic-ion Dual channels
Electron channel
Ion channel
Zn2+
SO42-
H2O
PSS
PEDOT
(g)
Voltage (V)
Capacity (mAh g-1)
1st
2000th
3000th
4000th
(h)
Capacity (mAh g-1)
Cycle number
Flat
60° Bending
120° Bending
180° Bending
Recovered

**Figure 4.** Mechanical resilience and electrochemical performance of water-rich QSE batteries. (a) Device during sequential cutting and (b) corresponding voltage after each cut. Reproduced with permission from Ref. [20]. Copyright 2018, Royal Society of Chemistry. (c) Photographs at flat, 60°, 90°, and 180° configurations; (d) galvanostatic charge–discharge profiles after 30 cycles at 50 mA $g^{-1}$; and (e) corresponding discharge capacities. Reproduced with permission from Ref. [117]. Copyright 2021, Wiley-VCH. (f) Schematics of Zn interfaces with SA and SA/EIDC gels; (g) charge–discharge profiles at selected cycles; and (h) pouch-cell capacity at different bending angles. Reproduced with permission from Ref. [118]. Copyright 2025, Elsevier.

**Dry and dynamically coordinated polymer electrolytes**

SPEs provide a route to leakage-free and flexible zinc batteries, but conventional PEO-based electrolytes are highly crystalline and have low room-temperature conductivity. L-serine-modified PEO/PVDF is one recent strategy for overcoming these limitations. L-serine reduces polymer crystallinity, introduces –OH coordination sites for $Zn^{2+}$, improves interfacial contact, and enhances mechanical robustness. Electrolytes without L-serine are designated PPZ, whereas those containing 20 mM L-serine are designated PPZS20. The optimized PPZS20 reached 0.102 mS $cm^{-1}$ and supported stable Zn plating/stripping for 1400 h. Figure 5a compares ion transport and deposition in PPZ and PPZS20. A PPZS20-based Zn||$MnO_2$ cell delivered 163–72 mAh $g^{-1}$ over 0.025–0.2 A $g^{-1}$ and recovered 101 mAh $g^{-1}$ when the current returned to 0.1 A $g^{-1}$ (Figure 5b). It retained 88 mAh $g^{-1}$ after 50 cycles (approximately 73% retention) and achieved an average Coulombic efficiency of 99.3% over 100 cycles (Figure 5c,d) [119]. Dynamic coordination polymers provide another route. In the PHP SPE, $Zn^{2+}$ coordinates with 2,6-bis((propylimino)methyl)-4-chlorophenol ligands, and rapid ligand exchange supports ion conduction and self-healing. The electrolyte is dry, stretchable, self-healing, and acid-degradable through its imine bonds. Figures 5e and 5f show the proposed intermolecular and intramolecular $Zn^{2+}$-exchange pathways, and Figure 5g shows the first four charge–discharge profiles at 0.5 C. The discharge capacity increased from 66.4 to 98.2 mAh $g^{-1}$, possibly as the electrode–electrolyte interface stabilized during cycling [120].

The central challenge for SPEs remains the trade-off between mechanical strength and ionic conductivity. Polymer chains must be mobile enough to transport $Zn^{2+}$ yet sufficiently robust to resist penetration and maintain contact. Many SPE full cells are also tested at elevated temperature, low cathode loading, or for few cycles. Future development should therefore emphasize room-temperature conductivity, thin-film processing, and intimate cathode–electrolyte contact.

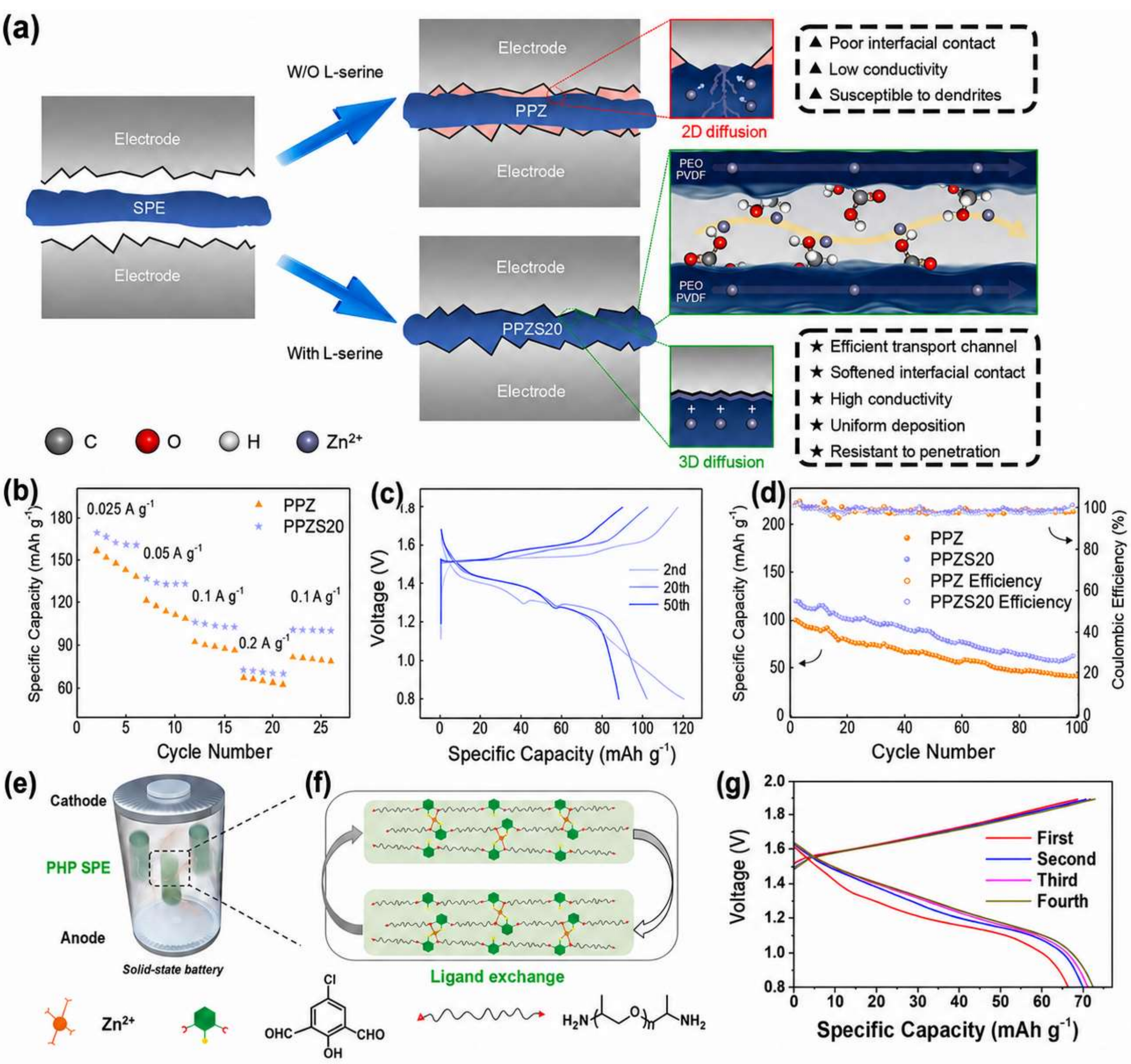

**Figure 5.** Dry polymer electrolytes and their cell performance. (a) Schematic comparison of PPZ and PPZS20; (b) rate performance of Zn||$MnO_2$ cells at 0.025, 0.05, 0.1, 0.2, and 0.1 A $g^{-1}$; (c) charge–discharge profiles of Zn|PPZS20|$MnO_2$ at selected cycles; and (d) cycling performance at 0.1 A $g^{-1}$. Reproduced with permission from Ref. [119]. Copyright 2026, Elsevier. (e) Solid-state battery based on the PHP SPE; (f) proposed $Zn^{2+}$ conduction mechanism; and (g) voltage–capacity profiles for the first four cycles. Reproduced with permission from Ref. [120]. Copyright 2021, American Chemical Society.

## Molecular-crystal and inorganic conductors

Crystallized eutectic conductors and succinonitrile (SN)-based molecular solids reveal important design rules for divalent-ion transport. In the $Zn^{2+}$-conducting electrolyte (ZCE), $TiO_2$ nanoparticles serve as nucleation seeds and Lewis-acidic sites. Adsorption of bis(trifluoromethanesulfonyl)imide ($TFSI^-$) on $TiO_2$ weakens $Zn^{2+}$–anion association and creates interfacial $Zn^{2+}$ percolation pathways. Thus, fast conduction can arise from engineered heterogeneous interfaces rather than bulk polymer motion alone. In Zn|ZCE|$V_2O_5$ cells, the $Zn^{2+}$ deintercalation peaks stabilized after the first cycle and agreed with the plateaus in the galvanostatic profiles (Figure 6a,b). After activation, the cell delivered 134.7 mAh $g^{-1}$ after 80 cycles, with an average Coulombic efficiency above 99.78% at 16 mA $g^{-1}$ and 30 °C (Figure 6c). SEM and AFM images after 80 cycles showed a smooth, dendrite-free Zn surface (Figure 6d,e). With a $Mo_6S_8$ cathode, the cell delivered 56.5 mAh $g^{-1}$ after activation and retained 94.3% of this capacity after 70 cycles at 16 mA $g^{-1}$ (Figure 6f) [60]. SN-based molecule-flexible solids further show that molecular rotation can assist $Zn^{2+}$ migration and that cation–anion pairing is a major transport limitation. In the earlier dry-state $ZnPS_3$ study, electronic conduction was negligible and $Zn^{2+}$ was identified as the mobile species, but the ionic conductivity remained too low for practical batteries [58]. Zeolite–Zn provides a more recent inorganic example. It uses low-cost zeolite, ion exchange, and cold pressing to form a $Zn^{2+}$-rich electrolyte sheet with high ionic conductivity, a high reported $Zn^{2+}$ transference number, corrosion suppression, and reduced vanadium dissolution. Figure 6g illustrates the proposed mechanism in Zn||$NH_4V_4O_{10}$ cells. CV established electrochemical reversibility (Figure 6h), and in situ EIS showed that charge-transfer resistance decreased and stabilized after approximately 25 cycles (Figure 6i). The Zeolite–Zn cell retained 84.9% of its capacity after 1000 cycles, whereas the $ZnSO_4$ control continued to decay within 800 cycles (Figure 6j) [53]. Table 3 compiles additional representative studies.

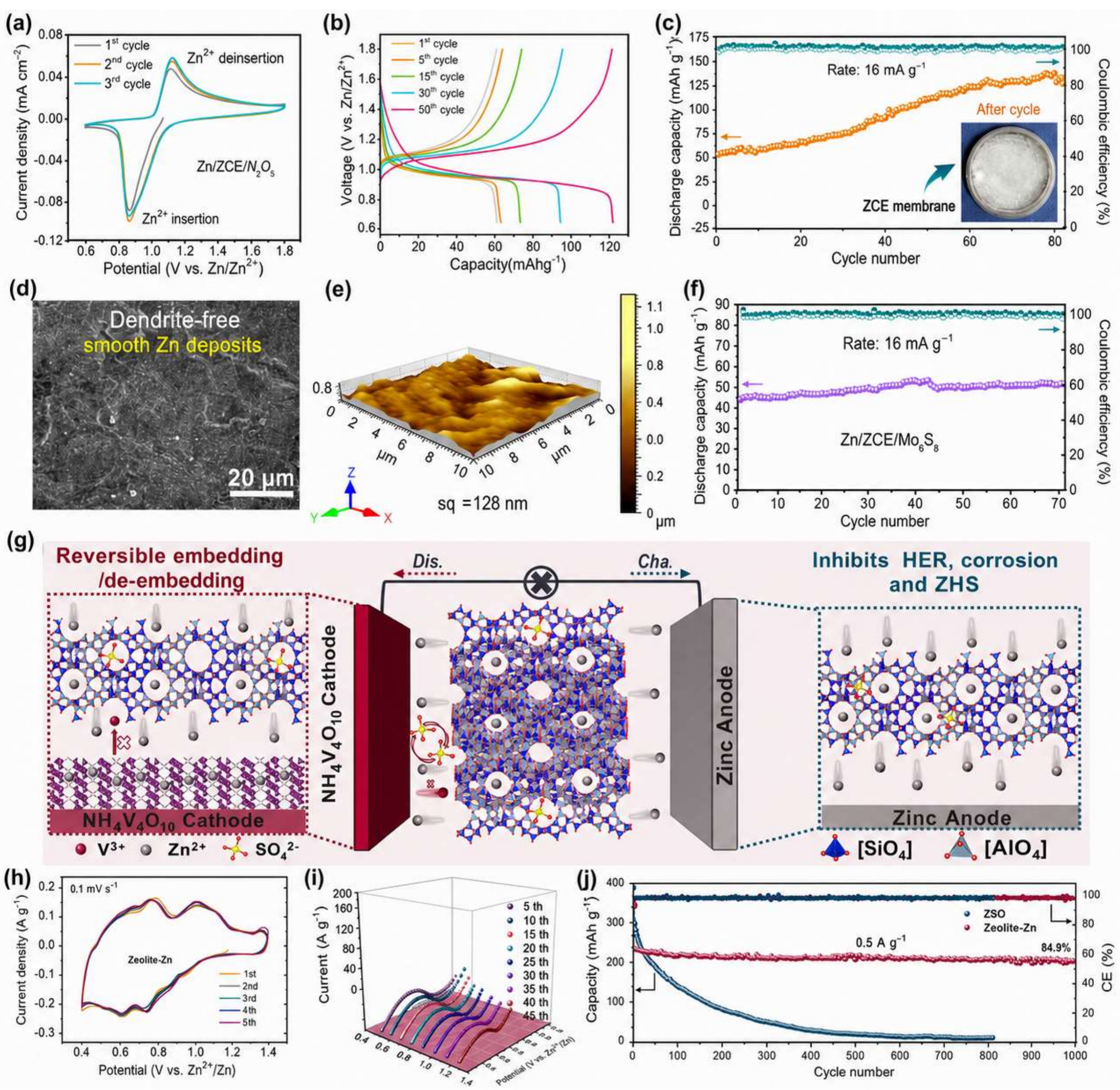

**Figure 6.** Molecular-crystal and inorganic $Zn^{2+}$ conductors. (a) CV curves of a Zn|ZCE|$V_2O_5$ cell at 0.2 mV s$^{-1}$; (b) charge–discharge profiles at 30 °C and 16 mA g$^{-1}$; (c) cycling performance and Coulombic efficiency; (d) SEM image and (e) AFM topography of the Zn anode after 80 cycles; and (f) cycling performance of a Zn|ZCE|$Mo_6S_8$ cell at 16 mA g$^{-1}$. Reproduced with permission from Ref. [60]. Copyright 2022, Wiley-VCH. (g) Proposed mechanism of Zeolite–Zn in a Zn||$NH_4V_4O_{10}$ cell; (h) CV curves; (i) in situ impedance evolution; and (j) long-term cycling in Zeolite–Zn and $ZnSO_4$ electrolytes at 0.5 A g$^{-1}$. Reproduced with permission from Ref. [53]. Copyright 2025, Wiley-VCH.

**Table 3.** Selected representative studies of QSEs and SEs for zinc batteries.

| No. | Electrolyte / type | Key composition | Representative cell performance | Main advantage | Main limitation |
|---|---|---|---|---|---|
| 1 | Chao et al., ZOV//Zn QSE-ZIB [121] | Fumed silica/$ZnSO_4$ gel with ZOV array cathode and Zn nanoflake array anode | 50 C rate capability; 2000 cycles at 20 C; ≈115 Wh $kg^{-1}$ and ≈5.1 kW $kg^{-1}$; >96% retention after 100 bending cycles | Excellent high-rate flexible device; electrode nanoarray reduces dendrite growth | Performance relies strongly on special electrode architecture, not only electrolyte; scalability of arrays is challenging |
| 2 | Li et al., gelatin-g-PAM/PAN hierarchical hydrogel electrolyte [20] | Gelatin-g-PAM hydrogel filled in PAN electrospun membrane | 306 mAh $g^{-1}$; 97% retention after 1000 cycles; areal energy density 6.18 mWh $cm^{-2}$ and power density 148.2 mW $cm^{-2}$ | Strong wearable demonstration; survives cutting, bending, hammering, puncturing, washing, fire, sewing | Still water-rich; practical encapsulation and calendar-life validation required |
| 3 | Qiu et al., ZCE crystallized eutectic electrolyte [60] | $Zn(TFSI)_2$-based deep eutectic system crystallized with $TiO_2$ nucleator | $Zn/V_2O_5$ solid cells delivered 134.7 mAh $g^{-1}$ within 80 cycles; average CE >99.78% | Mechanistic advance: interfacial $Zn^{2+}$ percolation and weakened ion association | Conductivity still lower than hydrogels; practical high-loading cell validation needed |
| 4 | Ma et al., ionic-liquid electrolyte and polymer SPE [122] | $[EMIM]BF_4$ + $Zn(BF_4)_2$; PVDF-HFP film filled with PEO/IL-based Zn salt | Zn plating/stripping >1500 h at 2 mA $cm^{-2}$; all-solid ZIB cycled 30,000 cycles at 2 A $g^{-1}$; stable from −20 to 70 °C | Hydrogen-free and dendrite-free strategy; strong abuse tolerance | Ionic-liquid cost, viscosity, and large-scale electrolyte supply must be considered |

| No. | Electrolyte / type | Key composition | Representative cell performance | Main advantage | Main limitation |
|---|---|---|---|---|---|
| 5 | Martinolich et al., inorganic $ZnPS_3$ conductor [58] | Layered insulating $ZnPS_3$ lattice | Fundamental conductivity study; no practical full ZIB | Important structure–property model for divalent conduction | Low conductivity and difficult sulfide processing limit near-term use |
| 6 | Iliescu et al., solvated all-Zn MOF conductor [110] | $Zn_3[(Zn_4Cl)_3(BTT)_8]_2 \cdot 5.2ZnCl_2$ with mobile extra-framework $Zn^{2+}$ cations, wetted with propylene carbonate | Ionic conductivity of $1.15 \times 10^{-4}$ S cm$^{-1}$ at 25 °C, activation energy of 0.317 eV, and electrochemical stability window above 2 V versus Zn | High room-temperature $Zn^{2+}$ mobility through open tetrazole coordination pockets in a fully exchanged, air-stable anionic framework | Multi-step post-synthetic cation exchange; moisture-sensitive when activated; demonstrated in pelletized conductivity cells without full-battery validation |
| 7 | Ling et al., in situ-polymerized eutectic electrolyte [62] | Ternary mixture of SN, N-methylacetamide, and $Zn(TFSI)_2$, cross-linked with ethoxylated trimethylolpropane triacrylate | 2.1 V discharge plateau; 81.2 mAh g$^{-1}$ with 81.2% retention after 2000 cycles at 5 C; pouch cell retained >80% after 900 cycles | Solvation regulation raises the full-cell discharge plateau and enables deep Zn cycling at high current density | Multi-component precursor requiring precise composition; thermally initiated in situ polymerization increases processing complexity and energy use |

| No. | Electrolyte / type | Key composition | Representative cell performance | Main advantage | Main limitation |
|---|---|---|---|---|---|
| 8 | Hui et al., polymer/MOF composite electrolyte (MOF-Fc@PM) [59] | Carboxyferrocene-functionalized cobalt MOF, $Zn(OTf)_2$, and a PEGDA/polyacrylamide copolymer matrix | Ionic conductivity of 1.52 mS cm$^{-1}$ and a reported $Zn^{2+}$ transference number of 0.83; 123.9 mAh g$^{-1}$ after 2000 cycles at 1.0 A g$^{-1}$ | Open $Co^{2+}$ sites anchor $OTf^{-}$ anions, while competitive PEGDA/PAM coordination accelerates $Zn^{2+}$ transfer | Multi-step synthesis involving defect-engineered MOF preparation, chemical functionalization, and radical copolymerization; cobalt and ferrocene increase cost |
| 9 | Yan et al., "plasticine-like" single-ion-conductive electrolyte (PLSE-In) [123] | PAN containing concentrated $(ZnCl_4)^{2-}$ complexes (approximately 43 m $ZnCl_2$) and a trace $InCl_3$ additive | Zn\|\|$I_2$: 450 mAh g$^{-1}$ at 0.2 A g$^{-1}$ and 89.28% retention after 1200 cycles at 1 A g$^{-1}$; Zn\|\|$Br_2$: 3500 cycles at 1.5 A g$^{-1}$ with 85.31% retention | Moldable contact, a reported $Zn^{2+}$ transference number of 0.90, broad electrochemical window, suppressed hydrogen evolution, and operation from −50 to 60 °C | High salt and $InCl_3$ cost; synthesis requires control of a metastable phase and prevention of salting-out |

| No. | Electrolyte / type | Key composition | Representative cell performance | Main advantage | Main limitation |
|---|---|---|---|---|---|
| 10 | Chen et al., in situ-polymerized amorphous SPE [4] | Poly(1,3-dioxolane) matrix formed in situ from 1,3-dioxolane and a Zn salt by Lewis-acid-initiated ring-opening polymerization | Room-temperature ionic conductivity of 19.6 mS cm⁻¹; encapsulation-free flexible devices operated for >30 days in air and tolerated >40 min of flame exposure | High reported conductivity and conformal, low-impedance interfaces produced by in situ polymerization | Polymerization is sensitive to moisture and protic impurities; residual monomer/mobile fractions and long-term depolymerization require quantification |

## 5. What changes inside an operating battery?

High bulk ionic conductivity does not by itself ensure effective $Zn^{2+}$ transport in an operating battery. Conductivity is generally measured with an electrolyte placed between blocking electrodes, whereas a working cell contains dynamically evolving Zn–electrolyte and cathode–electrolyte interfaces.

### Zn plating and stripping continuously reconstruct the interface

Zn plating is more than an ion-transfer process: it mechanically displaces the electrolyte contact, changes electrode thickness and roughness, and converts $Zn^{2+}$ to metallic Zn. Because Zn has a volumetric capacity of approximately 5855 mAh $cm^{-3}$, depositing 1 mAh $cm^{-2}$ produces about 1.7 µm of Zn; 3 and 10 mAh $cm^{-2}$ therefore correspond to thickness changes of approximately 5.1 and 17 µm, respectively [124]. These changes are substantial relative to SE films that may be only tens of micrometers thick. Uneven $Zn^{2+}$ flux concentrates current at surface protrusions during plating, creating localized deposits and mechanical stress. Preferential stripping can then produce voids, interfacial gaps, or electrically isolated Zn. A polymeric single-ion conductor illustrates the importance of stable mechanical and ionic contact: an ionic conductivity of 0.54 mS $cm^{-1}$ and a $Zn^{2+}$ transference number of 0.94 supported dendrite- and hydrogen-free Zn cycling for 2000 h [5]. The result reflects both favorable bulk transport and stabilization of the Zn interface, not conductivity alone.

**Interfacial contact can improve or deteriorate during cycling**

Solid–solid interfaces initially contain microscopic voids because neither Zn foil nor an SE surface is perfectly flat. Soft polymers and eutectic materials can deform to fill these voids, whereas rigid ceramics usually form discrete point contacts [125]. The local current density at these true contact points can therefore greatly exceed that calculated from the geometric electrode area. A recent hybrid metal-halide electrolyte illustrates the importance of contact retention. Its bulk conductivity was 0.29 mS cm$^{-1}$ at 25 °C, and X-ray computed tomography showed close Zn–electrolyte contact after 20 cycles. The resulting all-solid-state Zn/$I_2$ cell retained 234.5 mAh g$^{-1}$ after 200 cycles. The electrolyte pellet was approximately 430 μm thick and fabricated at about 20 MPa [126].

**Electrolyte thickness creates a direct conductivity penalty**

The bulk area-specific resistance scales linearly with electrolyte thickness, as shown in Equation (2).

$$R_{Bulk} \propto L \quad (2)$$

For an electrolyte with a conductivity of 10$^{-1}$ mS cm$^{-1}$, a 100 μm film contributes approximately 100 Ω cm$^2$, whereas a 430 μm pellet contributes approximately 430 Ω cm$^2$. Thinning is therefore one of the most direct ways to reduce cell resistance. Very thin electrolytes, however, are more vulnerable to pinholes, rough Zn deposits, mechanical penetration, drying, and handling damage. The relevant objective is the minimum defect-free thickness that maintains electronic insulation and interfacial integrity, not simply the thinnest possible membrane. A eutectic SE with a conductivity of 3.94 mS cm$^{-1}$ was incorporated as an approximately 20 μm film in pouch cells, corresponding to a calculated bulk contribution of only about 0.51 Ω cm$^2$. The conductivity measurement itself used a 3-mm-thick specimen, underscoring the need to distinguish characterization geometry from device geometry [62].

**Composite cathodes require simultaneous ionic and electronic networks**

A liquid electrolyte penetrates the pores of a conventional cathode, creating an extended three-dimensional reaction interface. A dry SE cannot automatically infiltrate these pores. If a cathode film is pressed against a solid-electrolyte pellet, only particles near the nominal planar interface may be electrochemically accessible.

A practical cathode must therefore contain (1) an electronically conductive network through the active material and carbon, (2) a continuous $Zn^{2+}$-conducting network through the electrolyte phase, (3) intimate active-material–electrolyte contact, and (4) sufficient porosity or free volume to accommodate structural changes. In situ polymerization is one strategy for meeting these requirements [62].

**Chemical and electrochemical stability are different requirements**

Chemical stability describes whether the electrolyte reacts spontaneously with Zn or the cathode under open-circuit conditions. Electrochemical stability describes whether it is oxidized or reduced when current and electrode polarization are applied. A material may satisfy one condition but fail the other. At the Zn interface, possible reactions include polymer or solvent reduction; corrosion and hydrogen evolution when residual water remains; formation of ZnO, $Zn(OH)_2$, carbonates, or basic Zn salts; dissolution or reconstruction of the electrolyte; and reactions between Zn and immobilized anionic groups. At the cathode interface, high potential may oxidize polymers, eutectic solvents or anions. Transition-metal cathodes may also catalyze electrolyte decomposition. Therefore, a voltage window measured by linear-sweep voltammetry on an inert electrode should not be treated as proof of long-term compatibility with a practical cathode.

For example, a Zn-compatible electrolyte containing an in situ-formed interphase maintained relatively stable interfacial resistance, whereas the charge-transfer resistance of a Zn|1 M $Zn(TFSI)_2$|Zn cell increased to 356 Ω after only 15 cycles. The increase was associated with hydrogen evolution and accumulation of ZnO, $Zn(OH)_2$, and carbonate-containing products [127].

**Interphase formation can be beneficial or blocking**

An interphase is not inherently harmful. Its effect depends on four properties: $Zn^{2+}$ conductivity, electronic insulation, chemical stability, and mechanical continuity. Interphase resistance must nevertheless be monitored over time.

A dense, electronically insulating but $Zn^{2+}$-conductive interphase can suppress electron transfer to water or other solvents while allowing Zn deposition beneath it. Conversely, an interphase dominated by poorly conducting oxides, hydroxides, or disconnected reaction products increases polarization. A hybrid-electrolyte-derived $Zn_5(CO_3)_2(OH)_6$-containing interphase exhibited ionic conductivities of 0.04–1.27 mS $cm^{-1}$ between −30 and 70 °C, showing that some Zn-containing interphases can transport $Zn^{2+}$ rather than merely passivate the metal [128]. Another study produced an approximately 20-nm-thick $ZnF_2$/ZnS-rich interphase that enabled Zn||Zn operation for more than 5000 h and a Coulombic efficiency of approximately 99.8% in Zn||Cu cells over 1500 cycles at 5 mA $cm^{-2}$ [129].

**Pressure is a transport variable, not merely an assembly detail**

Pressure improves contact by closing interfacial voids and increasing real contact area. It can also reduce particle–particle resistance in pressed electrolyte pellets and composite cathodes. Excessive pressure, however, may deform soft polymers, expel mobile solvent, fracture brittle ceramic particles, collapse cathode porosity, promote Zn penetration through defects, and make laboratory performance dependent on an impractical fixture. Pellet-formation pressure and operating stack pressure must therefore be reported separately. For example, the hybrid halide electrolyte was pelletized at approximately 20 MPa, whereas a solid-state eutectic pouch cell operated at approximately 101.3 kPa. These pressures serve different purposes and should not be treated as equivalent operating conditions [62,126].

## 6. What is practically credible?

Several general design principles can be extracted from these studies.

**Water activity must be controlled**

Water content and water activity describe different electrolyte properties. Water content is the total amount of water present, whereas water activity reflects its thermodynamic availability, or effective chemical potential, to participate in hydrogen evolution, Zn corrosion, and cathode dissolution [130]. An electrolyte may therefore contain substantial water at relatively low activity, or little water that remains highly reactive. Conceptually, the total water content can be divided into free and bound fractions.

Free water is relatively mobile, retains a bulk-like hydrogen-bonding environment, and can readily solvate ions or participate in side reactions. Bound water interacts strongly with ions or polar polymer groups and is generally less mobile and less chemically accessible. These categories are operational rather than absolute and depend on the measurement method.

Merely encapsulating an aqueous electrolyte within a polymer network can limit water movement and leakage, but it does not necessarily lower thermodynamic water activity. A genuine reduction in activity requires changes in water chemical potential, for example through high salt concentration, strong ion–water or polymer–water coordination, or eutectic interactions that disrupt the bulk water network. Claims of reduced water activity should therefore rely primarily on direct equilibrium measurements; spectroscopy, thermal analysis, and simulation should be used as mechanistic support. Lower water activity can suppress hydrogen evolution, corrosion, and cathode dissolution, but the accompanying loss of $Zn^{2+}$ mobility must also be quantified.

**$Zn^{2+}$ transference number matters**

High total ionic conductivity is insufficient when anion migration dominates. Electrolytes such as Zeolite–Zn and SA/EIDC report high $Zn^{2+}$ transference numbers, which can reduce concentration polarization and improve cycling stability. However, values derived from DC polarization in Zn symmetric cells should not be compared directly across gels, eutectics, and SEs because the measured current may not represent $Zn^{2+}$ transport through the bulk alone. Ion-concentration gradients, side reactions at the Zn–electrolyte interface, and changes in interfacial contact can affect both the initial and steady-state currents. Apparent transference numbers may therefore reflect cell construction and interfacial behavior as much as intrinsic ion transport. Meaningful comparison requires standardized electrode geometry, electrolyte thickness, polarization voltage, temperature, test duration, and impedance corrections for changes in bulk and interfacial resistance.

Cell-level area-specific resistance (ASR) is a more informative performance metric than conductivity alone. Equation (3) expresses the total ASR of a Zn battery containing an SE or QSE.

$$R_{ASR} = L / \sigma_{ion} + R_{Zn|SE} + R_{cathode|SE} \quad (3)$$

The total area-specific resistance comprises the bulk-electrolyte contribution and the two electrode–electrolyte interfacial contributions. In the bulk term, L denotes electrolyte thickness, and the denominator is the ionic conductivity. The Zn-side interfacial resistance can arise from poor physical contact, corrosion products, passivation layers, side reactions, or uneven Zn deposition. The cathode-side interfacial resistance reflects incomplete contact, limited electrolyte penetration, charge-transfer resistance, and slow interfacial ion transport. When thickness is expressed in cm and ionic conductivity in $S\ cm^{-1}$, the bulk term has units of $\Omega\ cm^2$; both interfacial terms must therefore also be area-normalized.

A material can have high ionic conductivity yet perform poorly if its interfacial resistance is large. Conversely, reducing electrolyte thickness lowers the bulk contribution but may not improve performance when either electrode interface dominates. Practical cells may also contain electronic-contact resistance, current-collector resistance, porous-electrode transport losses, and concentration polarization, as represented schematically by Equation (4).

$$R_{ASR} = L / \sigma_{ion} + R_{Zn|SE} + R_{cathode|SE} + R_{contact} + R_{porous\ electrode} \quad (4)$$

Thus, ionic conductivity alone is insufficient for comparing SEs; electrolyte thickness and both electrode–electrolyte interfaces must be evaluated under comparable conditions.

**Cathode loading and areal capacity**

The areal capacity is given by Equation (5).

$$Q_A = \frac{m_A q_{SP}}{1000} \quad (5)$$

In Equation (5), the specific capacity is expressed in mAh $g^{-1}$ and the active cathode loading in mg $cm^{-2}$. High mass loading with poor active-material utilization may be less meaningful than moderate loading with high areal capacity. Increasing loading produces thicker electrodes, longer $Zn^{2+}$ diffusion distances, greater tortuosity, and poorer electrolyte penetration. Electronic and ionic pathways may also become discontinuous. These problems are especially severe in SE- and QSE-based zinc batteries because the electrolyte cannot freely penetrate newly formed cracks or isolated pores.

A proton-selective coating enabled a high-loading $NaV_3O_8$ cathode to reach an areal capacity of 4.5 mAh $cm^{-2}$, while a two-layer pouch cell delivered 65.8 mAh and operated for 280 cycles. The coating also reduced dissolved V species from 3.7 to 0.6 mmol $L^{-1}$, showing that thick-electrode performance depends on both transport and cathode stability [131]. Another study tested $MnO_2$ and $Zn/MnO_2$ cathodes at loadings of 23.6 and 25.2 mg $cm^{-2}$, respectively, retaining 73% and 78% of their capacities after 5000 cycles with a protective cathode–electrolyte interphase [132].

**Zn utilization and the N/P capacity ratio**

Zn utilization is the fraction of available Zn capacity cycled during each discharge, as defined by Equation (6).

$$U_{Zn} = \frac{Q_{Cycled}}{Q_{Zn,total}} \times 100\% \quad (6)$$

Zn excess is commonly represented by the negative-to-positive capacity ratio, N/P, in Equation (7).

$$N/P = \frac{Q_{Zn,total}}{Q_{Cycled}} \tag{7}$$

$$\text{Thus, } U_{Zn} = \frac{1}{N/P} \times 100\% \tag{8}$$

when the cathode is fully utilized. Thick Zn foil can produce an apparently excellent cycle life because only a small fraction of the available Zn participates in each cycle. The large reservoir continuously replenishes Zn consumed by parasitic reactions or isolated as dead Zn.

For example, an ampere-hour pouch cell used 30 μm Zn with an areal capacity of 17.6 mAh $cm^{-2}$, a cathode areal capacity of 5.5 mAh $cm^{-2}$, and an N/P ratio of 3.2. It retained approximately 80% of its capacity after 390 cycles, corresponding to five months of operation [133]. A Zn–iodine pouch cell was tested at a reported Zn depth of discharge of 68.3% and retained 88.3% of its capacity after 5000 cycles, showing that high Zn utilization and long cycle life can be evaluated simultaneously [134].

**Cumulative capacity**

Cycling time alone does not define electrochemical throughput: a cell operated at low current and low capacity may show little cumulative Zn transfer. Cumulative areal capacity is therefore more informative and is calculated using Equation (9).

$$Q_{cum.} = NQ_{cycle} \tag{9}$$

In Equation (9), N is the number of plated or stripped half-cycles, and each half-cycle transfers the stated areal capacity. For example, 1000 half-cycles at 0.1 mAh $cm^{-2}$ correspond to only 0.10 Ah $cm^{-2}$. A test using 5 mAh $cm^{-2}$ per half-cycle would pass 50 times more Zn through the interface over the same number of half-cycles. The counting convention must be stated because some studies report one-direction plated capacity, whereas others report charge-plus-discharge throughput, which is twice as large.

A Zn||Zn cell containing an electrolyte additive delivered a cumulative capacity of 16.47 Ah $cm^{-2}$ at 20 mA $cm^{-2}$ and 1 mAh $cm^{-2}$. The same study tested a 30 μm Zn foil at 85.4% Zn depth of discharge for 150 h, providing a more demanding assessment than low-utilization symmetric-cell cycling [135].

**Electrolyte-to-capacity ratio**

The electrolyte-to-capacity ratio, E/C, describes the electrolyte inventory required per unit of cell capacity, as defined by Equation (10).

$$E\,/\,C = \frac{m_{electrolyte}}{Q_{Cell}} \tag{10}$$

The gravimetric E/C ratio has units of g $Ah^{-1}$. It may also be reported volumetrically as mL $Ah^{-1}$ or μL $mAh^{-1}$.

Large laboratory electrolyte volumes improve wetting, reduce concentration gradients, and replenish solvent consumed by side reactions. However, excess electrolyte lowers cell-level energy density and can conceal corrosion, water consumption, salt precipitation, and cathode dissolution. Reducing E/C creates more realistic but more demanding conditions because a small amount of decomposition can substantially change electrolyte concentration and water activity.

A practical ampere-hour Zn||$Zn_{0.25}V_2O_5 \cdot nH_2O$ pouch cell was operated at a cathode areal capacity of 5.5 mAh $cm^{-2}$, an N/P ratio of 3.2, and an E/C ratio of 9.3 g $Ah^{-1}$ (approximately 7 mL $Ah^{-1}$), under an external pressure of 0.1 MPa. It reached approximately 70 Wh $L^{-1}$ at the cell-component level and retained about 80% of its capacity after 390 cycles [133]. A more recent 1.27 Ah pouch cell used a cathode areal capacity of 7 mAh $cm^{-2}$ and an E/C ratio of 11.3 g $Ah^{-1}$, retaining 84% of its capacity after 120 cycles [136].

For gels and QSEs, E/C reporting requires additional care. Authors should separately report total gel mass, polymer-framework mass, salt mass, mobile water and solvent mass, electrolyte thickness, and electrolyte mass per unit cell capacity. Otherwise, a water-rich gel may appear to use "no liquid electrolyte" while still containing a large mobile-liquid inventory.

**Pouch-cell validation**

Pouch cells expose failure modes that may remain inconspicuous in small coin cells. Larger electrode areas amplify current nonuniformity, coating defects, edge reactions, and local drying. Multilayer cells also require uniform pressure, electrolyte distribution, and tab connection. Hydrogen evolution is especially important because gas accumulation separates electrode layers and causes pouch swelling. A meaningful pouch-cell demonstration should therefore report nominal and delivered capacity; number and dimensions of electrode layers; cathode loading and areal capacity; Zn thickness, N/P ratio, and Zn utilization; electrolyte volume and E/C ratio; applied stack pressure; packaging and tab masses included in energy-density calculations; and cell swelling or gas evolution.

A sulfolane-containing reverse-micelle electrolyte enabled an ampere-hour multilayer pouch cell to retain approximately 80% of its capacity after 390 cycles over five months under controlled N/P, E/C, and areal-capacity conditions [133]. An "open" 0.9 Ah gel-electrolyte pouch cell retained 84% of its capacity after 200 cycles under 370 kPa, illustrating how gas release, electrolyte retention, and mechanical pressure become integral to large-cell design [137].

**Calendar life and self-discharge**

Cycle life measures degradation caused by repeated charging and discharging, whereas calendar life measures deterioration over elapsed time, including storage and rest periods. Zinc batteries can lose capacity during open-circuit storage through Zn corrosion and hydrogen evolution, chemical self-discharge, cathode dissolution, spontaneous interphase growth, electrolyte evaporation or redistribution, and loss of electrode–electrolyte contact. The normalized self-discharge rate is defined by Equation (11).

$$r_{SD} = \frac{Q_{Before} - Q_{After}}{Q_{Before}\Delta t} \times 100\% \quad (11)$$

In Equation (11), $\Delta t$ may be expressed in days or months, and the numerator is the capacity loss measured over the specified interval. The Zn–iodine pouch cell cited above lost approximately 11.7% of its capacity per month while retaining 88.3% after 5000 cycles [134]. Another study incorporated twenty 24-hour aging periods into a limited-Zn full-cell protocol. At an N/P ratio of 2.5, the modified cell retained 83.37% of its capacity after 100 cycles, showing that rest periods can reveal corrosion losses hidden by continuous cycling [138].

## 7. Perspective: likely routes forward

The most realistic near-term pathway for SE- and QSE-based zinc batteries is not complete elimination of water or solvent, but controlled use of a limited mobile phase within a mechanically stable electrolyte. $Zn^{2+}$ has high charge density and coordinates strongly with water, organic ligands, and anions. Removing these species generally raises desolvation or hopping barriers, lowers room-temperature conductivity, and worsens solid–solid contact. Controlled-solvation and hybrid electrolytes therefore offer an intermediate strategy: retain sufficient mobile ligand for $Zn^{2+}$ dissociation and coordination exchange while suppressing water activity, leakage, corrosion, and cathode dissolution. Truly solvent-free $Zn^{2+}$ conductors remain an important longer-term target, but they require new crystal chemistry, defect engineering, and interface design.

**Near-term route: controlled solvation rather than indiscriminate drying**

Controlled-solvation strategies regulate the molecules and anions occupying the $Zn^{2+}$ coordination shell. The objective is not necessarily to minimize total water content, but to reduce reactive or weakly coordinated water while maintaining sufficiently rapid $Zn^{2+}$ transport. Relevant approaches include concentrated or dual-salt formulations, water–organic hybrid electrolytes, eutectic solvents, polymer-coordinated or bound-water systems, and immobilized ionic-liquid or solvent domains.

These formulations can lower water activity by strengthening water–salt, water–polymer, or water–cosolvent interactions. They may also promote anion participation in the $Zn^{2+}$ solvation shell and thereby alter interphase chemistry. Stronger coordination, however, must be balanced against slower ligand exchange and greater desolvation resistance.

Controlled-solvation electrolytes are also compatible with practical cell formats. A high-entropy-solvation electrolyte enabled an ampere-hour pouch cell to cycle for more than 250 cycles with Coulombic efficiency above 99.90% under a lean-electrolyte condition of 2.4 $\mu L\ mg^{-1}$ [139]. A low-concentration dual-salt electrolyte provided 15.1 $mS\ cm^{-1}$, a daily pouch-cell self-discharge rate of 0.13%, and 93% capacity retention after 900 cycles at 25 °C [140]. Similarly, a polymerized eutectic electrolyte reached 3.94 $mS\ cm^{-1}$ at 25 °C and supported Zn symmetric cells at 80% Zn utilization and 8 $mA\ cm^{-2}$ for 1700 h [62].

These results suggest that solvation control can improve interfacial reversibility without always requiring expensive water-in-salt concentrations. Nevertheless, such electrolytes must be evaluated for salt cost, viscosity, fluorinated-anion content, solvent toxicity, flammability, drying energy, and end-of-life separation.

## Hybrid electrolytes as the most practical engineering platform

Hybrid electrolytes combine liquid-like transport with solid-like dimensional stability. Representative structures include cross-linked polymers containing confined solvent, polymer–ceramic composites, polymerized eutectics, single-ion polymer domains, and porous frameworks filled with controlled liquid phases.

Their near-term advantage is primarily interfacial. A soft polymer or in situ-polymerized precursor can conform to rough Zn surfaces and penetrate porous cathodes, creating a larger effective contact area than a rigid ceramic pellet. This reduces the need for high stack pressure and supports continuous ionic pathways within thick composite cathodes. Hybrid matrices can also immobilize anions, regulate $Zn^{2+}$ flux, and accommodate the approximately 1.7 μm thickness change associated with plating 1 mAh $cm^{-2}$ of Zn.

A hybrid, however, should not automatically be described as all-solid-state. If its conductivity arises primarily from water, an ionic liquid, a eutectic solvent, or residual monomer, it remains solvent-containing even when it does not flow macroscopically. Essential reporting therefore includes water and organic-solvent mass fractions; residual monomer after polymerization; free, bound, and coordinated solvent fractions; evidence for any continuous mobile-liquid phase; and mass loss and conductivity under controlled humidity.

The likely near-term design is therefore a thin, conformal hybrid electrolyte containing the minimum mobile solvent needed to sustain $Zn^{2+}$ coordination exchange. Success should be judged by cell-level resistance, Zn utilization, and solvent retention—not simply by whether a membrane appears dry.

## Longer-term target: genuinely solvent-free $Zn^{2+}$ conductors

A genuinely solvent-free conductor should transport $Zn^{2+}$ through a solid framework without relying on mobile water, organic solvent, ionic-liquid domains, or residual monomer. Transport must instead proceed through vacancies, interstitial defects, coordination-site exchange, rotationally assisted hopping, or ordered framework channels.

This goal is particularly difficult because divalent $Zn^{2+}$ interacts strongly with surrounding anions through electrostatic and partly covalent interactions. Low mobile-carrier concentration and high migration barriers can therefore coexist even in Zn-rich frameworks. Total Zn content should not be confused with the concentration of mobile $Zn^{2+}$ defects.

Several reported solid conductors illustrate both progress and the remaining definitional problem:

1) A nominally dry $ZnPS_3$ conductor exhibited a migration barrier of approximately 0.3 eV and a conductivity of 2 mS cm$^{-1}$ at 30 °C. However, its proposed conduction mechanism involved bound water at grain surfaces, so humidity and residual-water dependence remain relevant [55].

2) Fluorinated mesoporous ZnSF reached 0.66 mS cm$^{-1}$ at 25 °C, but residual DMF molecules within the mesopores assisted surface transport. It is therefore an important solid host but not an unambiguous solvent-free conductor [48].

3) An in situ-polymerized poly(1,3-dioxolane) electrolyte was described as liquid-free and reached 19.6 mS cm$^{-1}$, but it was also described as "non-dry." Residual monomer and mobile molecular fractions must therefore be quantified before it is compared directly with dry ceramics or crystals [4].

4) An SN-based supramolecular crystal provided ordered $Zn^{2+}$ transport tunnels, a conductivity of approximately 0.6 mS cm$^{-1}$, and a reported $Zn^{2+}$ transference number of 0.97 at 25 °C. Although macroscopically crystalline, SN is a molecular coordination component, so phase mobility and long-term retention should still be reported [84].

5) A hybrid metal-halide crystal reached 0.29 mS cm$^{-1}$, a $Zn^{2+}$ migration barrier of 0.37 eV, and an electrochemical window of approximately 3.74 V. This material is closer to a framework-based solvent-free conductor, but its approximately 430 μm pellet thickness and 20 MPa fabrication pressure show that processing and interfacial resistance remain substantial challenges [126].

These examples show that the field has not yet established consistently whether high conductivity originates in the solid lattice, at grain surfaces, or within a residual molecular phase. The next scientific step is therefore not merely to report a higher conductivity, but to isolate and quantify its physical origin.

**A measurable research agenda**

Future progress should be assessed using standardized, cell-relevant benchmarks rather than conductivity or cycling time alone. Electrolyte studies should report water and solvent mass fractions, residual moisture, drying history, humidity dependence, mobile-liquid content, ionic conductivity, $Zn^{2+}$ transference number, activation energy, thickness, and area-specific resistance. Hybrid electrolytes should approach 1 mS $cm^{-1}$ at 25 °C. For genuinely solvent-free conductors, conductivity, thickness, and resistance must be specified as coupled targets: at 100 μm thickness, an ionic conductivity of at least 0.5 mS $cm^{-1}$ is required for a bulk ASR no greater than 20 Ω $cm^2$; if conductivity is only 0.1 mS $cm^{-1}$, thickness must not exceed 20 μm. A reported $Zn^{2+}$ transference number above 0.8 should be verified under a standardized protocol. Interfacial stability should be assessed by time-resolved impedance and limited-Zn testing, targeting Zn utilization above 50%, average Coulombic efficiency near 99.99%, and cumulative capacity above 1 Ah $cm^{-2}$. Full-cell validation should use cathode loadings above 10 mg $cm^{-2}$, areal capacities above 3 mAh $cm^{-2}$, N/P ratios below 3, and E/C ratios below 10 g $Ah^{-1}$. Ultimately, multilayer pouch cells of at least 0.5–1 Ah should retain more than 80% of their capacity for 500 cycles and demonstrate 6–12 months of calendar stability. These benchmarks should be evaluated alongside processing energy, solvent recovery, material cost, recyclability, and end-of-life treatment.

## 8. Conclusions

SEs and QSEs have significantly advanced the development of safer and more durable zinc batteries. Water-rich hydrogels currently offer the strongest combination of ionic conductivity, flexibility, and device-level performance for wearable and deformable electronics, but they do not eliminate hydrogen evolution, drying, freezing, or long-term water-management problems. SPEs and ionic-liquid-containing polymers offer more leakage-resistant alternatives, although conductivity and interfacial contact still require improvement. Crystallized eutectic conductors and molecule-flexible solids reveal important $Zn^{2+}$ transport principles, especially the need to reduce ion pairing and construct percolating interfacial pathways. Conventional ceramic and sulfide $Zn^{2+}$ conductors remain largely fundamental because of low room-temperature conductivity or difficult processing.

Ultimately, developing SEs and QSEs for zinc batteries is not simply a quest to eliminate liquid phases, but an exercise in managing coordination chemistry across bulk transport pathways and electrode interfaces. Rigid frameworks and polymer networks can lower water activity, suppress parasitic reactions, regulate Zn deposition, and broaden electrochemical stability, but practical viability depends on resolving trade-offs among ionic conductivity, interfacial resistance, and inactive mass. Commercially relevant progress therefore requires precise control of $Zn^{2+}$ solvation and minimization of inactive electrolyte inventory, rather than treating solidity as a performance metric in itself.

**Acknowledgements**

This work was supported by the JST Strategic International Collaborative Research Program (SICORP; JPMJSC25E2), the GIMRT Program of the Institute for Materials Research (IMR), Tohoku University (Proposal No. 202512-CRKKE-0212), and the AY2026 TUMUG Research Support Staff Program.